\documentclass[]{iopart}

\usepackage{graphicx}
\usepackage{dcolumn}
\usepackage{bm}
\pdfoutput=1
\begin{document}

\title{Stark deceleration of lithium hydride molecules}

\author{S. K. Tokunaga, J. M. Dyne, E. A. Hinds and M. R. Tarbutt}

\address{Centre for Cold Matter, Blackett Laboratory, Imperial
College London, Prince Consort Road,
London, SW7 2AZ, United Kingdom}
\ead{m.tarbutt@imperial.ac.uk}

\date{\today}

\begin{abstract}
We describe the production of cold, slow-moving LiH molecules. The molecules are produced in the ground state using laser ablation and supersonic expansion, and 68\% of the population is transferred to the rotationally excited state using narrowband radiation at the rotational frequency of 444\,GHz. The molecules are then decelerated from 420\,m/s to 53\,m/s using a 100 stage Stark decelerator. We demonstrate and compare two different deceleration modes, one where every stage is used for deceleration, and another where every third stage decelerates and the intervening stages are used to focus the molecules more effectively. We compare our experimental data to the results of simulations and find good agreement. These simulations include the velocity dependence of the detection efficiency and the probability of transitions between the weak-field seeking and strong-field seeking quantum states. Together, the experimental and simulated data provide information about the spatial extent of the source of molecules. We consider the prospects for future trapping and sympathetic cooling experiments.
\end{abstract}

\maketitle

\section{Introduction}

Cold polar molecules are desirable for a diverse range of applications in physics and chemistry \cite{Bethlem(1)03}. An extremely versatile way of producing fast-moving beams of molecules at temperatures of about 1\,K is the method of supersonic expansion \cite{Scoles,Campargue}. A gas cools to low temperature as it expands supersonically from high to low pressure and one has only to find a way of introducing the molecules of choice into this supersonically expanding gas. A variety of methods have been devised to achieve this including direct seeding, electric discharge, photo-dissociation, laser desorption and laser ablation. Though the molecules may initially be formed at very high temperatures, the supersonic expansion is extremely efficient at cooling them to the same temperature as the carrier gas, about 1\,K. The fast-moving molecules formed this way can then be decelerated to low speed using the Stark deceleration method \cite{Bethlem(1)99} or its optical \cite{Fulton(1)06} or magnetic \cite{Narevicius(1)08} analogues. In the Stark decelerator, molecules are slowed down by the force that a polarizable particle experiences in an electric field gradient. A spatially periodic electric field, set up by an array of identical electrodes, is switched on and off such that molecules with a range of initial positions and velocities are slowed down, bunched together and focussed through the apparatus. Once they are sufficiently slow, the molecules can be confined in a trap \cite{Bethlem(1)00} or storage ring \cite{Heiner(1)07} enabling spectroscopic and collisional studies of exceptional precision to be made.
%To date, NH$_3$, OH, NH and metastable CO have been decelerated and trapped, and Stark deceleration has been demonstrated for H$2$CO, SO$2$, and YbF.

In this paper we report the Stark deceleration of lithium hydride. Its large dipole moment of 5.88\,D \cite{Rothstein(1)69}, together with its low mass, make it relatively easy to control and make it an attractive choice for applications where dipole-dipole interactions are important. It has a simple, well-characterized structure, and because it has a large rotational constant and no lambda-doubling it is possible to produce a beam where most of the molecules are in a single quantum state. Unfortunately the alkali hydrides are difficult to produce in the gas-phase. We form cold LiH by laser ablation of a precursor target into a supersonically expanding gas \cite{Tokunaga(1)07}. This production mechanism differs from that of the other molecules that have been decelerated to low speed, and so our results demonstrate the great versatility of the technique. A similar production technique was used in experiments demonstrating the alternating gradient deceleration of YbF \cite{Tarbutt(1)04}.

To explore the very rich physics of ultracold dipolar gases, e.g. \cite{Goral(1)02}, the molecules produced by the deceleration technique need to be cooled further. One strategy is to sympathetically cool the trapped molecules using ultracold atoms \cite{Lara(1)06}. At present, there are very few calculations of the all-important elastic and inelastic collision cross-sections between atoms and molecules at low temperature. The LiH molecule is simple enough that very accurate calculations of these collision cross-sections are feasible. The work reported here is part of an effort to trap and sympathetically cool LiH molecules using ultracold alkali atoms. We note that the formation of ultracold LiH by the photoassociation of ultracold lithium and hydrogen is also being explored \cite{Juarros(1)06}.

\section{Experiment setup}

\begin{figure}
\centering
\includegraphics{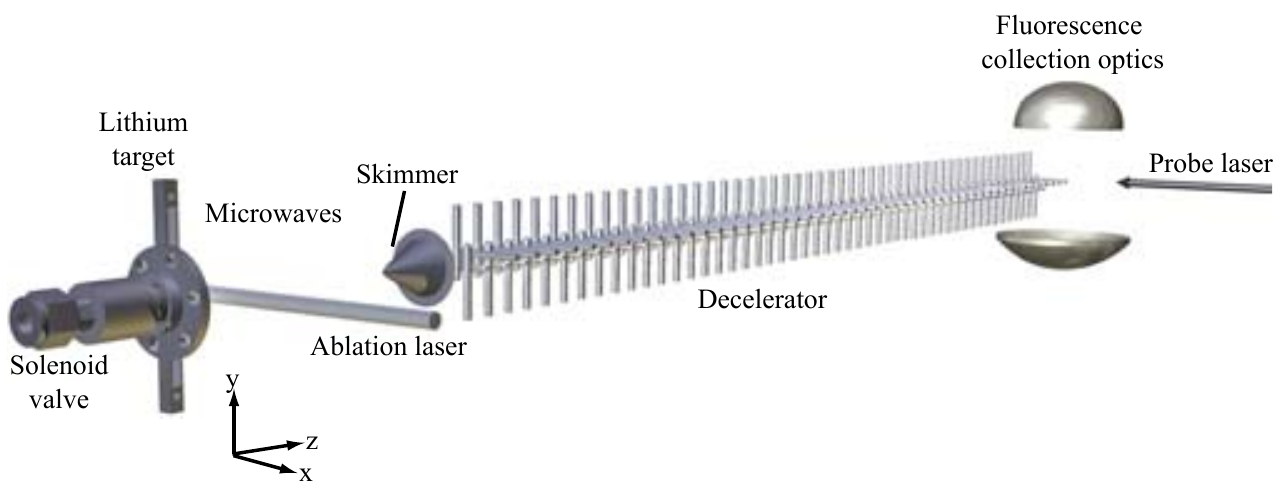}
\caption{Setup for Stark deceleration experiments
\label{Fig:Setup}}
\end{figure}

The experimental setup is shown in Fig.\,\ref{Fig:Setup}. Our methods for producing and detecting pulsed beams of cold LiH are detailed in \cite{Tokunaga(1)07}. The molecules are formed by laser ablation of a lithium target into a supersonically expanding gas mixture of Ar or Kr containing 2\% H$_{2}$. The gas expands through the 1\,mm diameter nozzle of a solenoid valve from a stagnation pressure of 4\,bar and temperature of 293\,K. The ablation laser is a Q-switched Nd:YAG laser with a wavelength of 1064\,nm, a pulse duration of $\approx 7$\,ns, an energy of $\approx 30$\,mJ and a repetition rate of 10\,Hz. Throughout this paper, we use a coordinate system with its origin at the centre of the valve nozzle, with the z-axis along the direction of the beamline and the y-axis vertical. The origin of time coincides with the firing of the ablation laser. The molecules pass through a skimmer at $z=93$\,mm, and are detected at $z=785$\,mm. The 2\,mm diameter opening of the skimmer separates the source chamber, where the pressure is $2 \times 10^{-5}$\,mbar during operation, from the vacuum chamber housing the decelerator where the pressure is maintained below $10^{-7}$\,mbar.

The Stark decelerator consists of 100 deceleration stages and is very similar in design to those used previously e.g. \cite{Bethlem(1)99}. The first stage is formed by a pair of oppositely charged 3\,mm diameter rods whose axes lie parallel to the $y-$axis and pass through the points $x=\pm2.5$\,mm, $z=135$\,mm. All odd-numbered stages are identical to the first and form an array with a $12$\,mm period along $z$. The even-numbered stages form a similar array, shifted by $6$\,mm along $z$ and rotated through 90$^{\circ}$ about the $z-$axis. Each of the 200 rods that make up the decelerator is mounted into one of four 16\,mm diameter support rods, according to its polarity and orientation. Four 20\,kV switches are used to switch between state one - odd numbered stages at high voltage, even stages grounded - and state two - even stages at high voltage, odd stages grounded. Two 20\,kV, $0.5\,\mu$F capacitors provide the charging currents and ensure that the voltage drop between the first and last high voltage pulses of the deceleration sequence is small. For high voltage conditioning, the switches and capacitors were removed, and the supplies connected to the electrodes via 1\,G$\Omega$ current-limiting resistors and robust floating nanoammeters \cite{Sauer(1)08}. The currents were recorded as the voltages were gradually increased. Because of electric field breakdown problems at $\pm 14$\,kV, we applied $\pm 10$\,kV in the deceleration experiments, rather than the design voltage of $\pm 20$\,kV.

The decelerator focusses molecules in weak-field seeking states. In our cold source, most of the molecules are in the strong-field seeking rotational ground state. We excite population into the weak-field seeking component of the first rotationally excited state, $(J=1,M=0)$, by driving the rotational transition in the region upstream of the skimmer. To produce narrow linewidth radiation at the 444\,GHz transition frequency we generated the fifth harmonic of an 88.8\,GHz Gunn oscillator which in turn was phase-locked to the fifth harmonic of a microwave synthesizer. The power obtained at the transition frequency was $\approx$60\,$\mu$W, and the linewidth was $\approx$15\,kHz. This radiation was coupled into free space via a pyramidal horn, collimated using a 50\,mm diameter plano-convex teflon lens placed one focal length (100\,mm) away from the apex of the horn, and passed into the vacuum chamber via a quartz window centred at $z=75$\,mm.

In the deceleration experiments, approximately 5\,mW of probe light from a frequency doubled cw Ti:Sapphire laser was used to detect $^7$LiH molecules in the $X\,^{1}\Sigma^{+}(v''=0,J''=1)$ state by laser-induced fluorescence on the $A\,^{1}\Sigma^{+}(v'=4)-X\,^{1}\Sigma^{+}(v''=0)$ R(1) transition at 367.2\,nm \cite{Tokunaga(1)07}. The collimated probe laser beam propagates along $x$, is linearly polarized along $z$, and is spatially defined by razor blades to fill a rectangular area $8$\,mm along $y$ and $3$\,mm along $z$. The fluorescence is collected on a photomultiplier tube and recorded as a function of time with a temporal resolution of $\approx$10\,$\mu$s. In taking data, we typically modulate, at a rate of 5\,Hz, between ``deceleration mode'' in which a deceleration timing sequence is applied, and ``dc mode'' where the voltages are applied to all the electrodes but not switched. This procedure eliminates the effect of a systematic drift in the source flux, allowing us to make reliable comparisons between decelerated and non-decelerated time-of-flight profiles.

\section{Population transfer}

\begin{figure}
\centering
\includegraphics{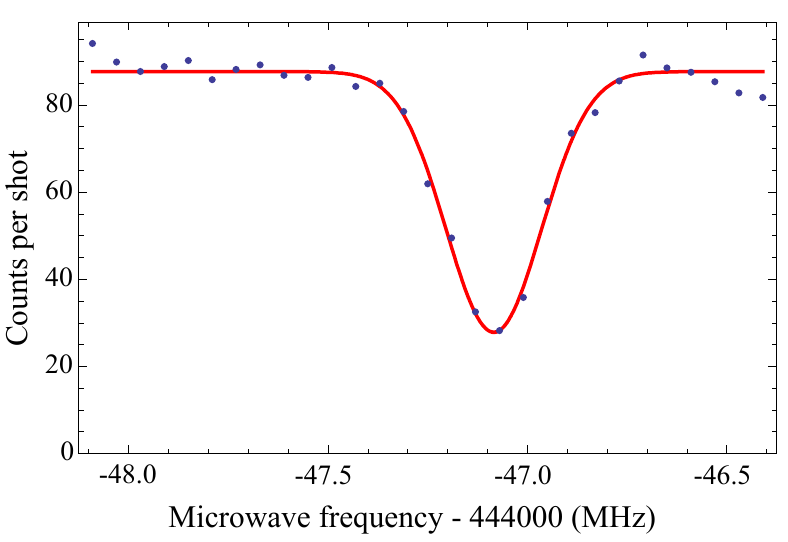}
\caption{Laser-induced fluorescence signal on the R(0) line as the microwave frequency is scanned through the rotational resonance. Points: experimental data. Line: Gaussian fit.
\label{Fig:Microwaves}}
\end{figure}

The first step of the experiment is to transfer population from the ground state to the first rotationally excited state. Figure \ref{Fig:Microwaves} shows the depletion of the ground state population as the microwave frequency is scanned across the rotational resonance, whose frequency is already well known \cite{Plummer(1)84,Bellini(1)95}. This data was taken without the decelerator in place. The ground state population was monitored by tuning the probe laser to the R(0) line, and the transition was driven using the maximum available microwave power. By fitting a Gaussian profile to this data we find a full width at half maximum (FWHM) of 281\,kHz, and a population transfer efficiency of 68\%. We emphasize that the rotational transition was driven in the source chamber, upstream of the skimmer. In this region we might expect the collision rate with the carrier gas atoms to be high enough to de-excite any population prepared in the $J=1$ state. The high efficiency that we observe shows that this collisional de-excitation is not a problem at a distance of $\sim$75\,mm from the source. We also note that we could not characterize the microwave beam in any way, other than by observing the effect on the molecules. We were not able to measure the power delivered to the molecules, and so relied on a previous calibration of the microwave source along with the literature values for the transmission of the lens and window materials. We could not measure the spatial profile of the microwave beam either.

A simple model of the transition lineshape, that includes the velocity distribution of the molecules, but excludes other details such as the spatial and angular distribution of the microwave power, gives the linewidth that we observe when we use our best estimates of the power and diameter of the microwave beam, 60\,$\mu$W and 3\,cm respectively. This same model gives a Lorentzian lineshape, whereas our data fits better to a Gaussian, and the model suggests that the population transfer should be 50\%, somewhat smaller than we observe. These differences are not surprising given the numerous details that are neglected in the simple model.

\section{Deceleration}

The Stark deceleration method has been discussed in detail elsewhere, and we shall follow this literature and use the usual terminology \cite{Bethlem(2)00,Meerakker(1)05}. We define the reduced position of a molecule, $\theta = \pi z/L$ (modulo $2\pi$), where $L=6$\,mm is the distance from one deceleration stage to the next. The $\theta=0$ position lies half way between one pair of electrodes and the next. The phase angle of a molecule, $\phi$, is its value of $\theta$ immediately after switching the decelerator from state one to state two. The synchronous phase angle, $\phi_{0}$, is the phase angle of the synchronous molecule, i.e. of the molecule whose value of $\phi$ remains constant throughout.

\begin{figure}
\centering
\includegraphics{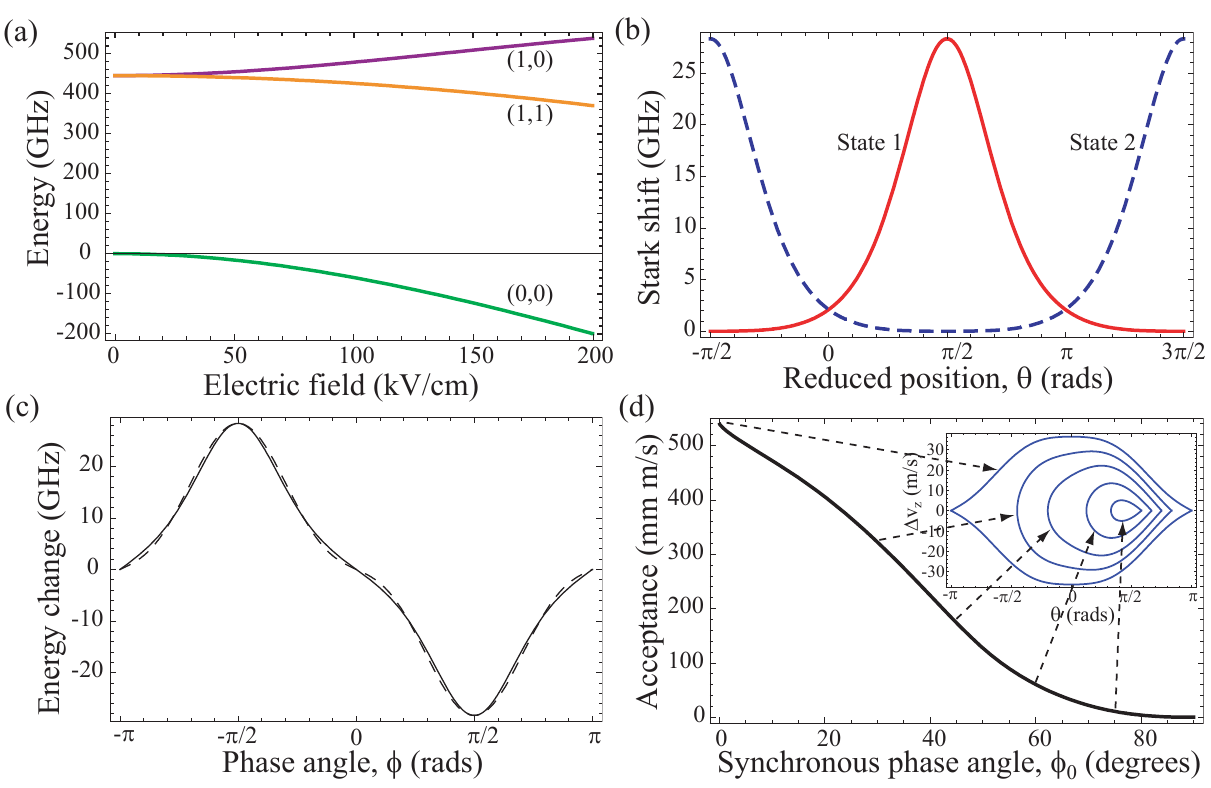}
\caption{(a) Stark shift of the low-lying states of $^{7}$LiH. States are labelled by the quantum numbers $(J,|M|)$. We decelerate molecules in the $(1,0)$ state. (b) Stark shift of the $(1,0)$ state as a function of the reduced position $\theta$ for applied voltages of $\pm10$\,kV. Both decelerator switch states are shown. (c) Solid line: change in kinetic energy per deceleration stage, $\Delta K$, as a function of the phase angle, $\phi$, for applied voltages of $\pm 10$\,kV. Dashed line: Eq.\,(\ref{Eq:DeltaKTwoTerms}) with $a=0.808$. (d) Longitudinal phase-space acceptance as a function of the synchronous phase angle for applied voltages of $\pm 10$\,kV, calculated using a one-dimensional model. The inset shows the accepted regions of phase space for $\phi_{0}=0^{\circ}, 30^{\circ}, 45^{\circ}, 60^{\circ}$ and $75^{\circ}$, obtained using the same model.
\label{Fig:Stark}}
\end{figure}

For many of the molecules decelerated previously, the Stark shift is approximately linear in the electric field magnitude and the change in the kinetic energy in one deceleration stage is well approximated by
\begin{equation}
\Delta K = -\Delta K_{{\rm max}} \sin\phi,\label{Eq:DeltaK}
\end{equation}
where $\Delta K_{{\rm max}}$ is the maximum possible change in the kinetic energy. For LiH however, this is a poor approximation because the Stark shift is far from linear at the field strengths relevant to the experiment. This can be seen in Fig.\,\ref{Fig:Stark}(a) which shows the Stark shifts of the three lowest-lying energy levels of $^{7}$LiH evaluated using a rigid rotor model of the molecule. We decelerate molecules in the $(1,0)$ state whose Stark shift  below 100\,kV/cm is approximately quadratic in the field strength. Figure \ref{Fig:Stark}(b) shows how the Stark shift in the two decelerator states depends on $\theta$. This is the potential in which the molecules move. Note that this potential is very flat in the regions close to the grounded electrodes because the electric field is small in this region and the Stark shift varies quadratically with the field. The change in kinetic energy in passing through one deceleration stage is a function of the phase angle and is found by taking the difference between the two curves shown in the figure. The result is shown by the solid line plotted in Fig.\,\ref{Fig:Stark}(c). It is evident from this graph that Eq.\,(\ref{Eq:DeltaK}) is inadequate in this case. Instead, we use the first two odd terms in a Fourier expansion, and write
\begin{equation}
\Delta K = \Delta K_{{\rm max}}(-a \sin(\phi)+(1-a)\sin(3\phi)), \label{Eq:DeltaKTwoTerms}
\end{equation}
where $a$ is a free parameter. In Fig.\,\ref{Fig:Stark}(c) we have plotted this equation (dashed line) for the best-fit value of $a=0.808$, showing that it provides a good approximation to the expected energy change.

To understand the longitudinal motion in a Stark decelerator, a good model is a series of travelling potential wells which are gradually slowed down \cite{Bethlem(2)00}. The synchronous molecule remains at the bottom of a potential well throughout, and molecules close to it in phase-space are trapped in that well. The depth of the well, and hence the number of trapped molecules, depends on $\phi_{0}$, becoming progressively shallower as $\phi_{0}$ increases from $0^{\circ}$ to $90^{\circ}$. Using Eq.\,(\ref{Eq:DeltaKTwoTerms}) and a model of the decelerator in which the longitudinal and transverse motions are not coupled, it is straightforward to obtain the equation of the separatrix, the curve that bounds all the molecules trapped in the well; this is Eq.\,(5) of \cite{Tarbutt(2)08}. Within this one-dimensional model, the longitudinal phase-space acceptance is obtained by calculating the area within the separatrix. For the parameters of our experiments, the result is shown as a function of $\phi_{0}$ in Fig.\,\ref{Fig:Stark}(d), together with plots of the separatrices for a few specific values of $\phi_{0}$. As outlined above, the acceptance falls with increasing $\phi_{0}$ meaning that the number of decelerated molecules will decrease as the degree of deceleration increases. While this result serves as a useful guide, the neglect of the coupling between longitudinal and transverse motions is usually a poor approximation. In particular the group of decelerated molecules tends to be more strongly focussed through the machine for larger values of $\phi_{0}$ \cite{Meerakker(1)06} and this partially counteracts the trend shown in Fig.\,\ref{Fig:Stark}(d).

\begin{figure}
\centering
\includegraphics{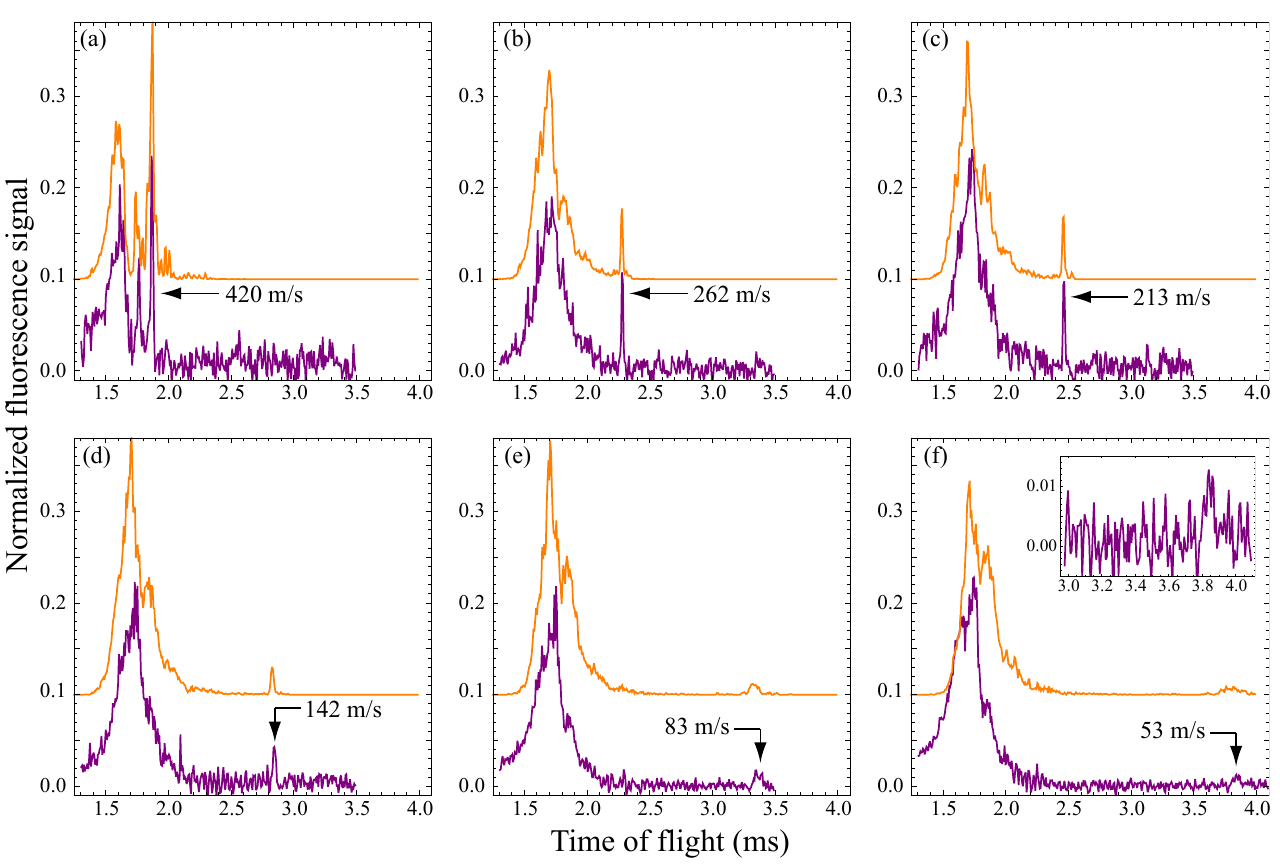}
\caption{Time-of-flight profiles for various values of the synchronous phase angle, showing the deceleration of molecules from an initial speed of 420\,m/s. For each dataset, a simulated profile is shown above the experimental data, offset for clarity. Each profile is normalized to the corresponding profile obtained in dc mode. The synchronous phase angles are (a) $0^{\circ}$, (b) $45^{\circ}$, (c) $51^{\circ}$, (d) $57^{\circ}$, (e) $60^{\circ}$, and (f) $60.9^{\circ}$. The magnified region in the inset to (f) shows the decelerated molecules more clearly.
\label{Fig:DecelSeries}}
\end{figure}

Figure \ref{Fig:DecelSeries} shows time-of-flight profiles recorded for various values of the synchronous phase angle, demonstrating deceleration from an initial speed of 420\,m/s. We have normalized each profile to the amplitude of the corresponding time-of-flight profile obtained in the dc mode of operation, in which the voltages are on but not switched. Because we modulate rapidly between the two modes of operation, the effect of slow drifts in the source flux are eliminated by this procedure. Above each experimental data set we show the prediction of a simulation, normalized in the same way. We discuss the experimental data first, returning to a comparison with these simulations later. For this data, Kr was used as the carrier gas and the LiH molecules were produced with a central speed of 439\,m/s and a translational temperature of 1.3\,K. Due to changes in the target condition, the mean speed of the beam entering the decelerator varies a little between datasets (from 431\,m/s to 448\,m/s), as does the translational temperature (from 1.1\,K to 1.9\,K). The initial speed of the synchronous molecule is determined by the electric field switching sequence and is always the same, 420\,m/s. The voltage applied to the decelerator's electrodes was nominally $\pm 10$\,kV. However, we found that the number of decelerated molecules was greatest when we applied a timing sequence calculated for $\pm 9.5$\,kV. We think this is partly due to an inaccuracy in the voltage measurement and partly due to the voltage drop between the first and last pulse of the sequence which was approximately 5\% and is not included in the algorithm that generates the timing sequence.

In Fig.\,\ref{Fig:DecelSeries}(a) the phase angle is zero, so there is no deceleration. The large narrow peak at 1.87\,ms corresponds to a group of molecules with a range of initial speeds centred at 420\,m/s. These molecules are all close in phase-space to the synchronous molecule and they remain trapped inside the travelling potential well. As a result they exit the decelerator in a compact bunch and form a narrow peak in the time-of-flight profile with very little signal on either side of this peak. There is also some contribution to this same peak from molecules that are confined in the potential wells on either side of the one occupied by the synchronous molecule. The smaller narrow peak that arrives earlier, at 1.78\,ms, is due to a group of molecules with speeds in the range 430-460\,m/s, whose phase-space positions at the entrance of the decelerator places them outside, but very close to, the separatrix. They form a thin band that follows the top-right portion of the separatrix of the main well, and the top-left portion of the next separatrix along. Their relative energy is slightly too large for them to be trapped in the well, so they continue to move ahead but spend a long time passing over the maxima of the potential wells. They diffuse less than they would if the potentials were absent and so they tend to be bunched at the detector. Because they spend the majority of their time near the maxima where the field is large and the focussing strong, they are also particularly well focussed through the decelerator. The large group of molecules having still earlier arrival times are ones that lie well outside the longitudinal phase-space acceptance. Their distribution of arrival times is not much affected by the decelerator.

In Fig.\,\ref{Fig:DecelSeries}(b), the phase angle is 45$^{\circ}$. Molecules are decelerated from 420\,m/s to 262\,m/s and appear as a narrow peak in the time-of-flight profile at 2.28\,ms. The rest of the profile, containing all the molecules outside of the phase-stable region, is similar to the profile obtained when the decelerator is not switched. As the phase angle is increased (Fig.\,\ref{Fig:DecelSeries}(c)-(f)), the final speed of the decelerated group decreases, and the peak in the time-of-flight moves to later arrival times. The amplitude of this peak also decreases, and this occurs for three reasons. Firstly, the longitudinal phase-space acceptance decreases with increasing phase angle and so the number of slow molecules decreases. Secondly, the slower the molecules the more they spread out longitudinally as they travel from the end of the decelerator to the detector. The peak therefore gets broader and smaller, this broadening being clearly visible in the data. Finally, the slower molecules spread out more in the transverse directions and they may miss the detection volume altogether. For our detector, this does not become significant until the speed is below 50\,m/s. In Fig.\,\ref{Fig:DecelSeries}(f) the decelerated bunch, shown more clearly in the inset, has a central speed of 53\,m/s and a temperature of approximately 10\,mK. These molecules are slow enough that they could be brought to rest at the centre of an electrostatic trap. Knowing the total detection efficiency \cite{Tokunaga(1)07} we estimate that there are only $\sim$10 LiH molecules per shot in this final decelerated bunch.

To understand our data more fully we simulated the experiments by calculating the trajectories of molecules through the entire apparatus from source to detector. The force on a molecule is derived from the Stark shift shown in Fig.\,\ref{Fig:Stark}(a) along with the electric field calculated using a finite element model of the decelerator. The timing sequences are the same as used in the experiments. The voltage drop between the first and last pulses of the sequence was not included. For each value of $\phi_{0}$, the time-of-flight profile formed by those molecules transmitted from source to detector (without crashing into any of the electrodes) is calculated, and then normalized to one calculated in dc mode. These are compared to the experimental data in Fig.\,\ref{Fig:DecelSeries} and are seen to reproduce that data rather well. They accurately predict the arrival time of the decelerated molecules, confirming that the final speed of the decelerated bunch is centred on the known final speed of the synchronous molecule. A small systematic discrepancy in the arrival times between simulation and experiment, that increases with decreasing speed, is consistent with the detector being 2.5\,mm further downstream than the nominal value; this is within the uncertainty of the detector placement. The amplitudes and widths of the decelerated peaks are also well matched by the simulations, the only exception being the 53\,m/s case where the peak seems to be narrower in the experimental data. Good agreement is also found for the amplitude of the un-decelerated fraction of the beam, although the shape of the un-decelerated profile is not accurately reproduced by the simulations at the larger phase angles. This is because the simulations assume a Gaussian distribution of initial velocities, while the profiles obtained in dc mode show that the true distributions are not Gaussian.

To obtain the good agreement between simulation and experiment that we see in Fig.\,\ref{Fig:DecelSeries}, we needed to include two additional effects, one which decreases the amplitude of the un-decelerated peak, and the other which increases the amplitude of the decelerated peak. The first effect is the possibility of transitions being driven to the $(J,|M|)=(1,1)$ state by the rapidly-changing orientation of the electric field during each switch. Because the molecules are strong-field seeking in this state, they are lost from the decelerator if this transition occurs. The probability for the transition depends on the angular frequency difference between the $(1,0)$ and $(1,1)$ states relative to the rate of change of the electric field orientation. Thus, the probability is highest for a molecule that is in the low field region between the grounded electrodes when the decelerator is switched. Referring to Fig.\,\ref{Fig:Stark}(b), such a molecule is close to $\theta=-\pi/2$ when the decelerator is switched from state 1 to state 2. Using the measured rise and fall times of 1\,$\mu$s and 0.5\,$\mu$s respectively, we find a significant transition probability in a narrow region centred on this position, but negligible probability elsewhere. Unless the decelerator is operated at low synchronous phase angles, the decelerated group of molecules are never in this region when the decelerator switches because their phase angles remain close to $\phi_{0}$ throughout. This is most clearly seen in the inset to Fig.\,\ref{Fig:Stark}(d) which shows that phase-stable molecules cannot reach the $\theta=-\pi/2$ position unless $\phi_{0}<30^{\circ}$. The molecules in the un-decelerated distribution explore all phase angles and so the probability of the transition becomes significant for them. Our simulations show that this failure to adiabatically follow the orientation of the electric field reduces the amplitude of the un-decelerated distribution by a factor of about two, while leaving the amplitude of the decelerated peak unchanged. For the $\phi_{0}=0^{\circ}$ data the situation is different, as explained above, and the amplitude of the phase-stable group is also reduced by this effect.

The second effect is the velocity dependence of the detection efficiency. At the probe intensity used, the laser induced fluorescence signal is not saturated and so the probability of excitation depends on the molecule-laser interaction time. Since the slower molecules interact with the laser for longer, they are more likely to be detected. To account for this, we used a simple three-level rate model \cite{Tarbutt(1)08, Wall(1)08} along with the known transition dipole moments \cite{Zemke(1)78}. While the un-decelerated fraction is hardly affected, the amplitude of the decelerated peak is considerably increased by this effect, e.g. by a factor of $\sim 1.8$ at 50\,m/s.

\begin{figure}
\centering
\includegraphics{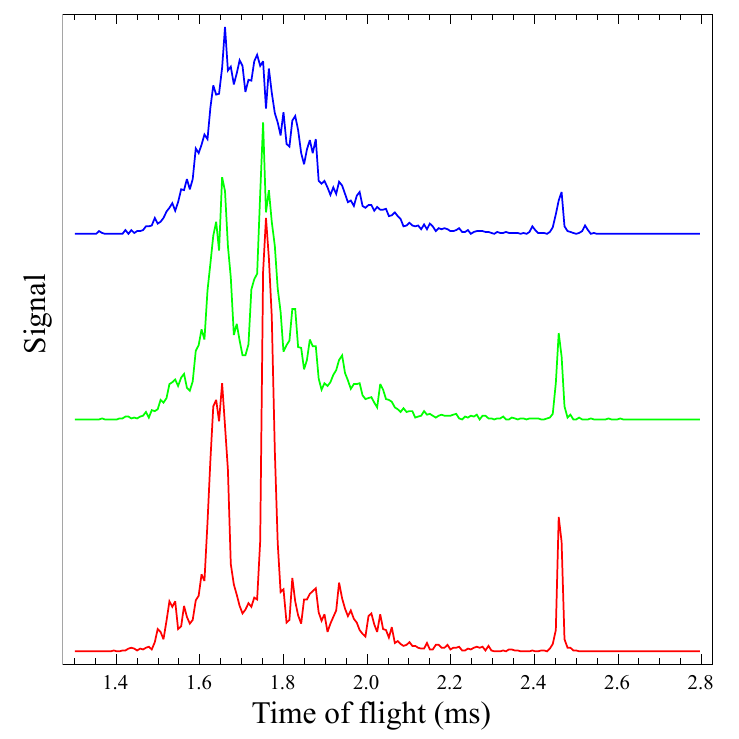}
\caption{Time-of-flight profiles resulting from simulations of a deceleration experiment with $\phi_{0}=51^{\circ}$, showing how narrow features in the profile are broadened out as the spatial extent of the source is increased. The widths are $\Delta z_{i}=2$\,mm (lower curve), 6\,mm (middle curve) and 12\,mm (upper curve).
\label{Fig:SourceSize}}
\end{figure}

In simulating the experiment, molecules are selected at random from an initial distribution in the source. For an accurate simulation, we need an accurate specification of the source, but our knowledge about the phase-space distribution of molecules in the source is quite limited. In the transverse directions we assume that the source is wide enough that the selection of molecules entering the decelerator is determined primarily by the skimmer and decelerator apertures. Then, the results are insensitive to the initial transverse distributions of positions and velocities. In the longitudinal direction, the time-of-flight profile measured with the decelerator off, or in dc mode, is an accurate reflection of the distribution of initial velocities, because the detector is very far away from the source compared to the source size. We use these measured distributions to fix the centre and the width of an initial Gaussian distribution of forward speeds in the simulations. This leaves only the longitudinal spread of initial positions undetermined. This spread is important because it determines the degree of correlation between the positions and speeds of molecules arriving at the decelerator.

To investigate the relevance of this parameter, we ran a set of simulations with various values of $\Delta z_{i}$, the FWHM of a Gaussian distribution of initial positions. For these simulations, we chose the same parameters as for the data in Fig.\,\ref{Fig:DecelSeries}(c). Figure \ref{Fig:SourceSize} shows the results obtained for $\Delta z_{i}=2,6$ and 12\,mm. We see that features that are narrow when the source is small are broadened out as the source is made larger, because a spread of initial positions translates into a spread of arrival times. For the 2\,mm source (lower curve) the most striking features are the two large peaks at 1.77 and 1.65\,ms, separated by the trough where there are very few molecules. These peaks are due to the untrapped dynamics of molecules that arrive at the decelerator approximately one and two periods ahead of the synchronous molecule. With increasing source size these peaks broaden out so that for the 12\,mm source (upper curve) the features have completely vanished. Comparing the simulated profiles obtained with various $\Delta z_{i}$ to the experimental data in Fig.\,\ref{Fig:DecelSeries}(c) we find the best agreement for $\Delta z_{i}=9$\,mm, and we conclude that this is the size of our source. We used this value for all the simulations shown if Fig.\,\ref{Fig:DecelSeries}.

The phase-space acceptance of a Stark decelerator is greatly affected by coupling between the longitudinal and transverse motions \cite{Meerakker(1)06, Sawyer(1)08, Tarbutt(3)08}. This coupling can lead to unstable regions of phase-space within the region that would otherwise be stable, and it can reduce the phase-space acceptance by a large factor, particularly for low values of $\phi_{0}$. Although in a one-dimensional model deceleration at low $\phi_{0}$ offers high longitudinal acceptance (Fig.\,\ref{Fig:Stark}(d)), this advantage is lost when coupling between transverse and longitudinal motion is present \cite{Scharfenberg(1)08}. This has been investigated experimentally using OH radicals, and we find it to be of great importance in our experiments too. Indeed, the reduction in the transverse focussing when $\phi_{0}$ is small is even more pronounced in LiH because of the quadratic nature of the Stark shift at low field. Coupling is inherent to the Stark decelerator because the same elements are used to both decelerate and focus the beam. In an elegant strategy that reduces the coupling while retaining the same electrode structure, the synchronous molecule travels a distance $s L$ between switches, rather than $L$, where $s$ is an odd integer greater than 1 \cite{Meerakker(1)05, Meerakker(1)06, Scharfenberg(1)08}. In this way, the role of the majority of the electrodes is to focus the beam, with only a subset being used for deceleration. Although the decelerator then needs to be longer, the number of decelerated molecules can be increased.

\begin{figure}
\centering
\includegraphics{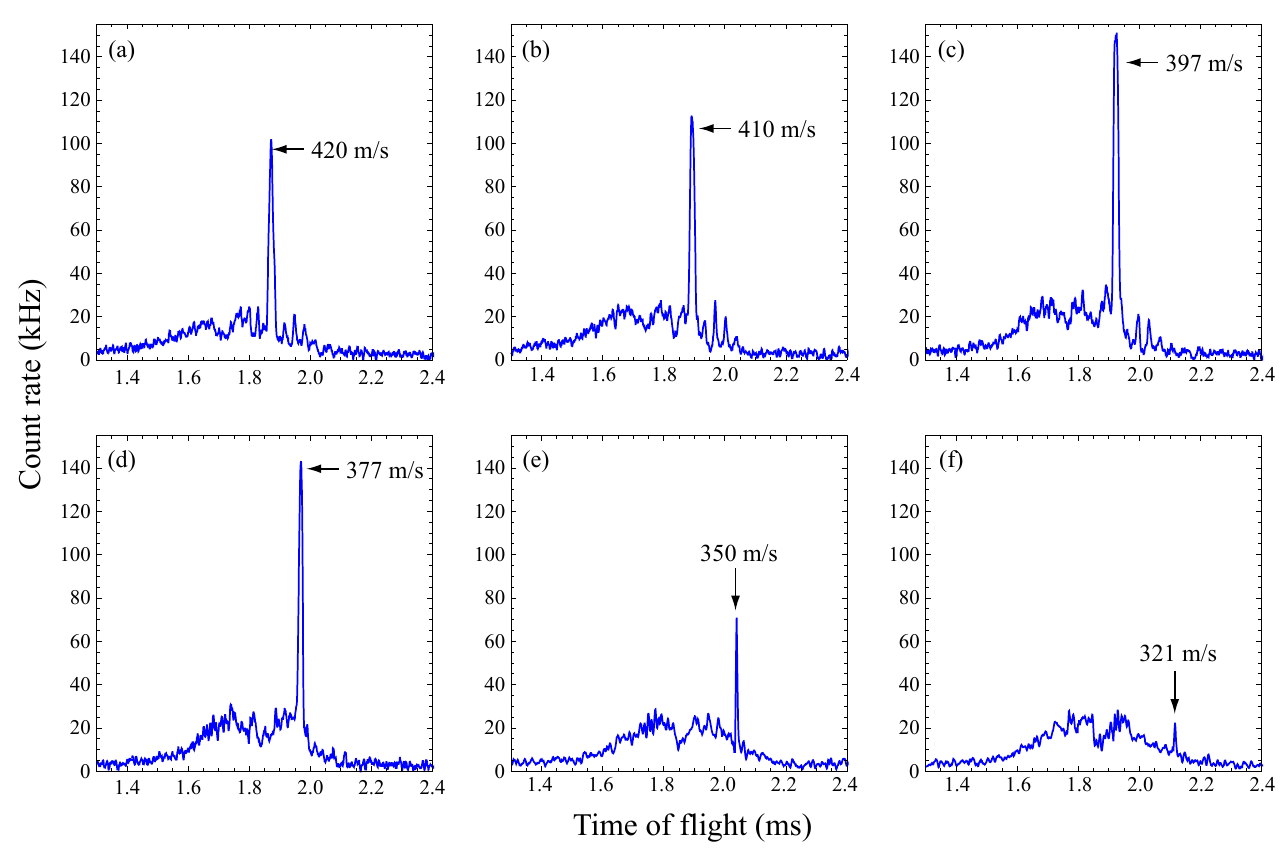}
\caption{Experimental time-of-flight profiles showing deceleration in the $s=3$ mode. The initial speed of the synchronous molecule is 420\,m/s and the synchronous phase angles are (a) $0^{\circ}$, (b) $15^{\circ}$, (c) $30^{\circ}$, (d) $45^{\circ}$, (e) $60^{\circ}$, and (f) $75^{\circ}$.
\label{Fig:Decels3}}
\end{figure}

Figure \ref{Fig:Decels3} shows the deceleration of LiH measured using the $s=3$ mode of operation at various values of $\phi_{0}$. The applied voltage was $\pm 10$\,kV and the initial speed of the synchronous molecule was 420\,m/s. The beam entering the decelerator had a central speed of 424\,m/s and a temperature of 0.7\,K. In these experiments the source flux was rather stable and we have plotted the detected photon count rate directly instead of normalizing to data obtained in dc mode. For reference, the profile has an amplitude of approximately 70\,kHz in dc mode. In Fig.\,\ref{Fig:Decels3}(a) the phase angle is zero and molecules with speeds centred around 420\,m/s are transported through the decelerator confined to the travelling potential well, appearing at the detector as a large, narrow peak at 1.87\,ms. The amplitude of this peak is about 7 times larger than the corresponding peak seen for $s=1$ operation (Fig.\,\ref{Fig:DecelSeries}(a)). In addition, there is a set of far smaller narrow peaks in the profile spaced by intervals of approximately 40\,$\mu$s. These are due to the dynamics of molecules that are outside the phase-stable region and so not trapped in the travelling potential wells. Although they are not phase-stable their motion is nevertheless influenced by the switched electric field. Those molecules with initial phase-space positions close to (but outside) the separatrix are influenced particularly strongly. Their motion is modulated considerably as they travel through the potential wells, they spend more time near the maxima of those wells, and this leads to the small peaks in the time-of-flight profile.

As $\phi_{0}$ is increased, the phase-stable group is decelerated, appearing at the detector at later times. For values of $\phi_{0}$ up to $45^{\circ}$ the amplitude of the decelerated bunch does not change much, the variation being consistent with the fluctuations in source flux between datasets. Since the longitudinal acceptance decreases with increasing $\phi_{0}$ we would expect this amplitude to decrease. We suggest two reasons, in addition to changes in the flux, why that is not observed. Firstly, as discussed above, the number of molecules in the phase-stable group is reduced by transitions driven by the switched electric fields, and this is effective only when $\phi_{0}<30^{\circ}$. Secondly, although the dependence of the focussing on $\phi_{0}$ is far weaker in the $s=3$ mode, it is still the case that focussing is a little less effective at lower $\phi_{0}$. When $\phi_{0} = 45^{\circ}$ and $s=3$, the amplitude of the decelerated peak is about 20 times larger than the one obtained for $s=1$ at the same $\phi_{0}$. Of course, the energy removed is reduced by a factor of 3 and so the decelerator needs to be 3 times longer to obtain the same final speed at the same phase angle. For large $\phi_{0}$ the amplitude of the decelerated peak drops because the longitudinal acceptance is reduced. The smaller peaks due to the untrapped molecules are still observed in all the datasets though they are far less pronounced at higher values of $\phi_{0}$. Note that the overall effect of transitions to the $(1,1)$ state is less important in the $s=3$ data because the transitions are induced by the switching field and the number of such switches is reduced by a factor of 3. The velocity dependence of the detection efficiency also plays only a minor role because the relative change in the velocity is quite small.

\section{Conclusions}

In this paper, we have demonstrated the state preparation and subsequent Stark deceleration of LiH molecules. The molecules are produced in the rotational ground state by laser ablation seeding of a supersonic expansion, and then transferred to the rotationally excited state with an efficiency of up to 68\%. Starting from an initial speed of 420\,m/s, the molecules have been decelerated to 53\,m/s, removing 98.5\% of the kinetic energy in preparation for subsequent trapping experiments. Our measurements are in good agreement with the results of simulations. These simulations include the possibility that molecules change state as the field switches and they include the speed dependence of the detection efficiency. By comparing experimental and simulated time of flight profiles we also obtain an estimate of the longitudinal extent of the LiH distribution produced in our source.

LiH is an attractive molecule for many applications in the physics and chemistry of cold molecules. The number of slow LiH molecules produced in our apparatus is very small at present and this number will need to be increased enormously for those applications to be realized. The primary limitation is the very low flux produced by the source and so we are now investigating alternative production mechanisms. In addition to an improved source, there are several other potential improvements to be made to the apparatus. In the present experiments, population was transferred from the ground state to all three components of the $J=1$ state, but only the $(1,0)$ state is useful in the experiment. By driving the transition in a small static electric field, this component can be populated selectively and, assuming the same transfer efficiency, the flux of slow molecules increased by a factor of 3. We note that our decelerator electronics are designed for $\pm 20$\,kV, double the voltages used in the present experiments. Following careful conditioning we expect to be able to operate the decelerator at this higher voltage. Our simulations show that this will increase the number of decelerated molecules by a factor of 25. At this higher voltage it is possible to decelerate to rest using $\phi_{0}=23^{\circ}$ if the initial speed is 420\,m/s, or $\phi_{0}=17^{\circ}$ if the initial speed is reduced to 350\,m/s by using Xe as the carrier gas. Operation at such low phase angles does not bring much advantage in the $s=1$ mode of operation due to the coupling of the longitudinal and transverse motions \cite{Scharfenberg(1)08}. At this higher voltage the $s=3$ mode can also be used to bring molecules to rest, requiring phase angles of $\phi_{0}=55^{\circ}$ and $42^{\circ}$ for these same two choices of initial speed. For low final velocities, operation in the $s=3$ mode ceases to be a useful strategy for increasing the number of slow molecules because of the problem of excessive focussing in the decelerator at low speed, and so dedicated focussing and slowing elements may then be useful \cite{Sawyer(1)08}. If the required increase in the number of slow LiH molecules can be obtained, we will trap the molecules together with ultracold alkali atoms with a view to sympathetic cooling of the molecules.

\ack

We thank Arzhang Ardavan and Philippe Goy for lending us the microwave apparatus and teaching us to use it. We are indebted to Jony Hudson for the software used to control the experiment and to Ben Sauer for his advice on numerous technical matters. This work was supported by the UK EPSRC and STFC, by the Royal Society, and the the ESF EuroQUAM programme as part of the CoPoMol project.

\end{document}